\documentclass[runningheads]{llncs}
\usepackage{longtable} 
\usepackage{graphicx,subfigure}
\usepackage{subfigure}
\usepackage{latexsym}
\usepackage{amsmath}
\usepackage{amssymb}
\usepackage{enumerate}
\usepackage{alltt}
\usepackage{url}
\usepackage[export]{adjustbox}
\usepackage{xcolor,colortbl}
\usepackage{multirow}
\usepackage{flushend}
\usepackage{proof}
\usepackage{array,multirow}
\usepackage{enumitem}
\usepackage{comment}

\usepackage{algorithm}
\usepackage{algpseudocode}
\usepackage{subfig}
\usepackage{booktabs} 
\usepackage[title]{appendix}

\begin{document}
\frontmatter
\setcounter{page}{1}

\mainmatter

\newcommand{\ctlX}{\mathsf{X}}        
\newcommand{\ctlF}{\mathsf{F}}        
\newcommand{\ctlG}{\mathsf{G}}            
\newcommand{\U}{\mathcal{U}}
\newcommand{\R}{\mathcal{R}}
\newcommand{\W}{\mathcal{W}}
\newcommand{\X}{\bigcirc}        
\newcommand{\F}{\Diamond}        
\newcommand{\G}{\Box}            
\newcommand{\False}{\textit{false}}     
\newcommand{\True}{\textit{true}}      
\newcommand{\true}{\textit{true}}      
\newcommand{\Implies}{\rightarrow}  
\newcommand{\Iff}{\leftrightarrow}  

\newcommand{\ApproxMC}{\#\textsc{Approx}}

\newcommand{\answer}[1]{{\color{darkgreen}[#1]}}
\newcommand{\red}[1]{{\color{black}{#1}}}
\newcommand{\blue}[1]{{\color{black}{#1}}}
\definecolor{darkgreen}{rgb}{0,0.65,0}

\newcommand{\fontsizeformula} {\fontsize{8}{8}}
\algnewcommand{\LineComment}[1]{\State \(\triangleright\) #1}
\algrenewcommand\algorithmicindent{1.0em}

\newcommand{\OurTool}{\textsc{ACoRe}}

\setcounter{page}{1}
\addtocontents{toc}{\protect\section*{First Topic}}

\title{ACoRe: Automated Goal-Conflict Resolution}
\author{Luiz Carvalho$^{1}$, Renzo Degiovanni$^{1}$, Matías Brizzio$^{2,3}$, Maxime Cordy$^{1}$, Nazareno Aguirre$^{4}$, Yves Le Traon$^{1}$, Mike Papadakis$^{1}$}
\institute{$^{1}$ SnT, University of Luxembourg, Luxembourg \\ \{firstname.surname\}@uni.lu \\ $^{2}$ IMDEA Software Institute, Spain; $^{3}$ Universidad Politécnica de Madrid, Spain \\ matias.brizzio@imdea.org \\ $^{4}$ Universidad Nacional de Río Cuarto and CONICET, Argentina \\ naguirre@dc.exa.unrc.edu.ar}
\authorrunning{Carvalho et al.}
\maketitle

\begin{abstract}
System goals are the statements that, in the context of software requirements specification, capture how the software should behave. Many times, the understanding of stakeholders on what the system should do, as captured in the goals, can lead to different problems, from clearly contradicting goals, to more subtle situations in which the satisfaction of some goals inhibits the satisfaction of others. These latter issues, called \emph{goal divergences}, are the subject of \emph{goal conflict analysis}, which consists of identifying, assessing, and resolving divergences, as part of a more general activity known as goal refinement. 

While there exist techniques that, when requirements are expressed formally, can automatically identify and assess goal conflicts, there is currently no automated approach to support engineers in \emph{resolving} identified divergences. In this paper, we present {\OurTool}, the first approach that automatically proposes potential resolutions to goal conflicts, in requirements specifications formally captured using linear-time temporal logic. {\OurTool} systematically explores syntactic modifications of the conflicting specifications, aiming at obtaining resolutions that disable previously identified conflicts, while preserving specification consistency. {\OurTool} integrates modern multi-objective search algorithms (in particular, NSGA-III, WBGA, and AMOSA) to produce resolutions that maintain coherence with the original conflicting specification, by searching for specifications that are either \emph{syntactically} or \emph{semantically} similar to the original specification.

We assess {\OurTool} on 25 requirements specifications taken from the literature. We show that {\OurTool} can successfully produce various conflict resolutions for each of the analyzed case studies, including resolutions that resemble specification repairs manually provided as part of conflict analyses. 
\end{abstract}
\section{Introduction}
\label{sec:introduction}

Many software defects that come out during software development originate from incorrect understandings of what the software being developed should do~\cite{MendezWKFMVC17}. These kinds of defects are known to be among the most costly to fix, and thus it is widely acknowledged that software development methodologies must involve phases that deal with the elicitation, understanding, and precise specification of \emph{software requirements}. Among the various approaches to systematize this requirements phase, the so-called \emph{goal-oriented requirements engineering} (GORE) methodologies~\cite{Dardenne+1993,VanLamsweerde2009} provide techniques that organize the modeling and analysis of software requirements around the notion of \emph{system goal}. Goals are prescriptive statements that capture how the software to be developed should behave, and in GORE methodologies are subject to various activities, including goal decomposition, refinement, and the assignment of goals 
\cite{Alrajeh+2009,Dardenne+1993,Degiovanni+2014,Letier2002,VanLamsweerde2009,vanLamsweerde1998}. 

The characterization of requirements as formally specified system goals enables tasks that can reveal flaws in the requirements. Formally specified goals allow for the analysis and identification of \emph{goal divergences}, situations in which the satisfaction of some goals inhibits the satisfaction of others~\cite{CailliauVanLamsweerde2015,Degiovanni+2018}. These divergences arise as a consequence of \emph{goal conflicts}. A \emph{conflict} is a condition whose satisfaction makes the goals inconsistent. Conflicts are dealt with through \emph{goal-conflict analysis}~\cite{vanLamsweerdeLetier2000}, which comprises three main stages: \emph{(i)} the \emph{identification} stage, which involves the identification of conflicts between goals; \emph{(ii)} the \emph{assessment} stage, aiming at evaluating and prioritizing the identified conflicts according to their likelihood and severity; and \emph{(iii)}, the \emph{resolution stage}, where conflicts are resolved by providing appropriate countermeasures and, consequently, transforming the goal model, guided by the criticality level. 

Goal conflict analysis has been the subject of different automated techniques to assist engineers, especially in the conflict identification and assessment phases~\cite{Degiovanni+2018,Degiovanni+2016,LuoWSYZC21,vanLamsweerde1998}. However, no automated technique has been proposed for dealing with goal conflict \emph{resolution}. In this paper, we present {\OurTool}, the first automated approach that deals with the goal-conflict resolution stage. {\OurTool} takes as input a set of goals formally expressed in Linear-Time Temporal Logic (LTL)~\cite{MannaPnueli1992}, together with previously identified conflicts, also given as LTL formulas. It then searches for candidate \emph{resolutions}, i.e., syntactic modifications to the goals that remain consistent with each other, while disabling the identified conflicts. More precisely, {\OurTool} employs modern search-based algorithms to efficiently explore syntactic variants of the goals, guided by a syntactic and semantic similarity with the original goals, as well as with the inhibition of the identified conflicts. This search guidance is implemented as (multi-objective) fitness functions, using Levenshtein edit distance~\cite{10.1145/3300148} for syntactic similarity, and approximated LTL model counting~\cite{DBLP:journals/corr/abs-2105-12595} for semantic similarity. {\OurTool} exploits this fitness function to search for candidate resolutions, using various alternative search algorithms, namely a Weight-Based Genetic Algorithm (WBGA)~\cite{DBLP:books/mit/H1992}, a Non-dominated Sorted Genetic Algorithm (NSGA-III)~\cite{DebJain2014}, an Archived Multi-Objective Simulated Annealing search (AMOSA)~\cite{DBLP:journals/tec/BandyopadhyaySMD08}, and an unguided search approach, mainly used as a baseline in our experimental evaluations.

Our experimental evaluation considers 25 requirements specifications taken from the literature, for which goal conflicts are automatically computed~\cite{Degiovanni+2018}. The results show that {\OurTool} is able to successfully produce various conflict resolutions for each of the analysed case studies, including resolutions that resemble specification repairs manually provided as part of conflict analyses. In this assessment, we measured their similarity concerning the ground-truth, i.e., to the manually written repairs, when available. The genetic algorithms are able to resemble 3 out of 8 repairs in the ground truth.
Moreover, the results show that {\OurTool} generates more non-dominated resolutions (their finesses are not subsumed by other repairs in the output set) when adopting genetic algorithms (NSGA-III or WBGA), compared to AMOSA or unguided search, favoring genetic multi-objective search over other approaches. 

\section{Linear-Time Temporal Logic}
\label{sec:background}

\subsection{\red{Language Formalism}}
\label{sec:ltl}

Linear-Time Temporal Logic (LTL) is a logical formalism widely used to specify reactive systems~\cite{MannaPnueli1992}. \red{In addition, GORE} methodologies (e.g. KAOS) have also adopted LTL to formally express requirements~\cite{VanLamsweerde2009} and taken advantage of the powerful automatic analysis techniques associated \blue{with} LTL to improve the quality of their specifications (e.g., to identify inconsistencies~\cite{Degiovanni+2018b}). 

\begin{definition}[LTL Syntax] 
Let $AP$ be a set of propositional variables. LTL formulas are inductively defined using the standard logical connectives, and the temporal operators $\X$ (next) and $\U$ (until), as follows:
\begin{enumerate}[label=(\alph*)]
\item constants $\True$ and $\False$ are LTL formulas; 
\item every $p \in AP$ is an LTL formula;
\item if $\varphi$ and $\psi$ are LTL formulas, \blue{then} $\neg \varphi$, $\varphi \vee \psi$, $\X \varphi$ and $\varphi \U \psi$ \blue{are also LTL formulas}.
\end{enumerate}
\end{definition}

LTL formulas are interpreted over infinite traces of the form $\sigma = s_0\ s_1 \ldots$, where each $s_i$ is a propositional valuation on $2^{AP}$ (i.e., $\sigma \in 2^{AP^\omega}$).  
\begin{definition}[LTL Semantic] 
We say that trace $\sigma = s_0, s_1, \ldots$ satisfies a formula $\varphi$, written $ \sigma \models \varphi$, if and only if $\varphi$ holds at the initial state of the trace, i.e. $(\sigma, 0) \models \varphi$. The last notion is inductively defined on the shape of $\varphi$ as follows:
\begin{enumerate}[label=(\alph*)]
\item $(\sigma, i) \models p \Leftrightarrow p \in s_i$
\item $(\sigma, i) \models (\phi \vee \psi) \Leftrightarrow (\sigma, i) \models \phi \text{ or }  (\sigma, i) \models \psi$
\item $(\sigma, i) \models \neg \phi \Leftrightarrow (\sigma, i) \not\models \phi$
\item $(\sigma,i) \models \X \phi \Leftrightarrow   (\sigma, i+1) \models \phi$
\item $(\sigma, i) \models (\phi\ \U\ \psi) \Leftrightarrow   \exists_{k \geq 0}: (\sigma, k) \models \psi \text{ and } 
\forall_{0 \leq j < k} : (\sigma, j) \models \phi$	
\end{enumerate}
\end{definition}

Intuitively, formulas with no temporal operator are evaluated in the first state of the trace. Formula $\X \varphi$ is true at position $i$,  iff $\varphi$ is true in position $i+1$. 
Formula $\varphi \U\ \psi$ is true in $\sigma$ iff formula $\varphi$ holds at every position until $\psi$ holds. 

\begin{definition}[Satisfiability] 
An LTL formula $\varphi$ is said \emph{satisfiable} (SAT) iff there exists at least one trace satisfying $\varphi$. 
\end{definition}

We also consider other typical connectives and operators, such as, $\land$, $\G$ (always), $\F$ (eventually) and $\W$ (weak-until), that are defined in terms of the basic ones. 
That is, $\phi \land \psi \equiv \neg(\neg \phi \vee \neg \psi)$, $\F \phi \equiv \true \U \phi$, $\G \phi \equiv \neg \F \neg \phi$, and $\phi \W \psi \equiv (\G \phi) \lor (\phi \U \psi)$.

\subsection{Model Counting}
\label{sec:model-counting}
The \emph{model counting} problem consists of calculating the number of models that satisfy a formula. 
Since the models of LTL formulas are infinite traces, it is often the case that analysis is restricted to a class of canonical \emph{finite} representation of \emph{infinite} traces, such as lasso traces or tree models. Notably, this is the case in bounded model checking for instance~\cite{Biere+1999}.

\begin{definition}[Lasso Trace]
A lasso trace $\sigma$ is of the form $\sigma = s_0 \ldots \ s_i (s_{i+1}$ $\ldots s_k)^\omega$, where the states $s_0 \ldots s_k$ conform the \emph{base} of the trace, and the loop from state $s_k$ to state $s_{i+1}$ is the part of the trace that is repeated infinitely many times. \end{definition}

For example, an LTL formula $\G(p \lor q)$ is satisfiable, and one satisfying lasso trace is $\sigma_1 = \{p\}; \{p,q\}^\omega$, wherein the first state $p$ holds, and from the second state both $p$ and $q$ are valid forever. Notice that the base in the lasso trace $\sigma_1$ is the sequence containing both states $\{p\}; \{p,q\}$, while the state $\{p, q\}$ is the sequence in the loop part. 

\begin{definition}[LTL Model Counting] 
\label{def:MC}
Given an LTL formula $\varphi$ and a bound $k$, the (bounded) \emph{model counting} problem consists in computing how many lasso traces of at most $k$ states exist for $\varphi$.
We denote this as $\#(\varphi,k)$.
\end{definition}

Since existing approaches for computing the exact number of lasso traces are ineffective~\cite{Finkbeiner+2014}, Brizzio et. al~\cite{DBLP:journals/corr/abs-2105-12595}  recently developed a novel  model counting approach that approximates the number (of prefixes) of lasso traces satisfying an LTL formula. Intuitively, instead of counting the number of lasso traces of length $k$,
the approach of Brizzio et. al~\cite{DBLP:journals/corr/abs-2105-12595}  aims at approximating the number of bases of length $k$ corresponding to some satisfying lasso trace. 

\begin{definition}[Approximate LTL Model Counting] 
\label{def:approxMC}
Given an LTL formula $\varphi$ and a bound $k$, the approach of Brizzio et. al~\cite{DBLP:journals/corr/abs-2105-12595} approximates the number of bases $w = s_0 \ldots s_k$, such that for some $i$, the lasso trace $\sigma = s_0 \ldots \ (s_i \ldots s_k)^\omega$ satisfies $\varphi$ (notice that prefix $w$ is the base of $\sigma$). 
We denote $\ApproxMC(\varphi,k)$ to the number computed by this approximation.
\end{definition}
{\OurTool} uses $\ApproxMC$ model counting to compute the semantic similarity between the original specification and the candidate goal-conflict resolutions. 
\section{The Goal-Conflict Resolution Problem}
\label{sec:GORE}
Goal-Oriented Requirements Engineering (GORE)~\cite{VanLamsweerde2009} drives the requirements process in software development from the definition of high-level goals that state how the system to be developed should behave.
Particularly, goals are prescriptive statements that the system should achieve within a given domain. The domain properties are descriptive statements that capture the domain of the problem world. 
Typically, GORE methodologies use a logical formalism to specify the expected system behavior, e.g., \red{KAOS uses Linear-Time Temporal Logic for specifying requirements~\cite{VanLamsweerde2009}}. 
\blue{In this context,} a \textit{conflict} essentially represents a condition whose occurrence results in the loss of satisfaction of the goals, i.e., that makes the goals \textit{diverge}~\cite{vanLamsweerde1998,vanLamsweerdeLetier1998}. Formally, it can be defined as follows.

\begin{definition}[Goal Conflicts]
\label{def:goal-conflicts}
Let $G = \{G_1,\ldots,G_n\}$ be a set of goals, and $Dom$ be a set of domain properties, \red{all written in LTL}. 
Goals in $G$ are said to diverge if and only if there exists at least one \textit{Boundary Condition (BC)}, such that the following conditions hold:
\begin{itemize}
\item \textit{logical inconsistency:} 
$\{Dom ,BC,\textstyle\bigwedge\limits_{1\leq i \leq n} G_i \} \models \False$
\item \textit{minimality:} for each $1\leq i \leq n$, 
$\{ Dom, BC, \textstyle\bigwedge\limits_{j \neq i} G_j \} \not \models \False$
\item \textit{non-triviality:}
$BC  \neq \neg (G_1 \land \ldots \land G_n)$
\end{itemize}
\end{definition} 
Intuitively, a BC captures a particular combination of circumstances in which the goals cannot be satisfied. The first condition establishes that, when $BC$ holds, the conjunction of goals $\{G_1,\ldots, G_n\}$ becomes inconsistent. 
The second condition states that, if any of the goals are disregarded, then consistency is recovered. 
The third condition prohibits a boundary condition to be simply the negation of the goals. 
Also, the minimality condition prohibits that $BC$ be equals to $\False$ (it has to be consistent with the domain $Dom$).

Goal-conflict analysis~\cite{VanLamsweerde2009,vanLamsweerde1998} deals with these issues, through three main stages: (1) The goal-conflicts identification phase consists in generating boundary conditions that characterize divergences in the specification; (2) The assessment stage consists in assessing and prioritizing the identified conflicts according to their likelihood and severity; (3) The resolution stage consists in resolving the identified conflicts by providing appropriate countermeasures. 
\red{Let us consider the following examples found in our empirical evaluation and commonly presented in related works.} 


\begin{example}[Mine Pump Controller - MPC]
\label{exa:MPC}
Consider the Mine Pump Controller (MPC) widely used in related works that deal with formal requirements and reactive systems~\cite{Degiovanni+2018,Kramer+1983}. 
The MPC \blue{describes} a system that is in charge of activating or deactivating a pump ($p$) to remove the water from the mine, \blue{in the} presence of possible dangerous scenarios. 
The MP controller monitors environmental magnitudes related to the presence of methane ($m$) and the high level of water ($h$) in the mine. Maintaining a high level of water for a while may produce flooding in the mine, while the methane may cause an explosion \red{when} the pump is switched on. 
Hence, the specification for the MPC is as follows:
\begin{align*}
& Dom: \G ( ( p \wedge \X(p) ) \Implies \X(\X(\neg h))  \hspace*{.45cm}
G_1:  \G (m \Implies \X(\neg p)) \hspace*{.45cm}
G_2: \G (h \Implies \X(p))
\end{align*}


Domain property $Dom$ describes the impact \red{into the environment of} switching on the pump ($p$).
\red{For instance}, \red{when} the pump is kept on for 2 unit times, then the water will decrease and the level will not be high ($\neg h$).
Goal $G_1$ expresses that the pump should be off when methane is detected in the mine. Goal $G_2$ indicates that the pump should be on when the level of water is high.

Notice that this specification is consistent, for instance, in cases in which the level of water never exceeds the high threshold. 
However, approaches for goal-conflict identification, such as the one of Degiovanni et al.~\cite{Degiovanni+2018}, can detect a conflict between goals in this specification. 

The identified goal-conflict describes a divergence situation 
in cases in which the level of water is high and methane is present at the same time in the environment.
Switching off the pump to satisfy $G_1$ will result in a violation of goal $G_2$; while switching on the pump to satisfy $G_2$ will violate $G_1$. 
This divergence situation clearly evidence a conflict between goals $G_1$ and $G_2$ that is captured by a boundary condition such $BC = \F (h \wedge m)$.
\end{example}

In the work of Letier et al.~\cite{Letier2001} two resolutions were manually proposed that precisely describe what should be the software behaviour in cases where the divergence situation is reached.
The first resolution proposes to refine goal $G_2$, by weakening it, requiring to switch on the pump only when the level of water is high and no methane is present in the environment.  
\begin{example}[Resolution 1 - MPC]
\begin{align*}
& Dom: \G ( ( p \wedge \X(p) ) \Implies \X(\X(\neg h)) ) \\
& G_1: \G (m \Implies \X(\neg p)) \hspace*{.45cm} 
G_2': \G (h \land \neg m \Implies \X(p))
\end{align*}
\end{example}

With a similar analysis, the second resolution proposes to weaken $G_1$, requiring switching off the pump when methane is present and the level of water is not high. 
\begin{example}[Resolution 2 - MPC]
\begin{align*}
& Dom: \G ( ( p \wedge \X(p) ) \Implies \X(\X(\neg h)) ) \\
& G_1':  \G (m \land \neg h \Implies \X(\neg p)) \hspace*{.45cm} G_2: \G (h \Implies \X(p))
\end{align*}
\end{example}

The \textit{resolution} stage aims at removing the identified goal-conflicts from the specification, for which it is necessary to modify the current specification formulation. 
This may require weakening or strengthening the existing goals, or even removing some and adding new ones.

\begin{definition}[Goal-Conflict Resolution]
\label{def:conflict-resolution}
Let $G = \{G_1,\ldots,G_n\}$, $Dom$, and $BC$ be the set of goals, the domain properties, and an identified boundary condition, respectively \red{written in LTL}. 
Let $M: S_1 \times S_2 \mapsto [0,1]$ and $\epsilon \in [0,1]$ be a similarity metric between two specifications and a threshold, respectively. 
We say that a resolution $R = \{R_1,\ldots,R_m\}$ resolves goal-conflict $BC$, if and only if, the following conditions hold:
\begin{itemize}
\item \textit{consistency:} $\{Dom, R \} \not\models \False$
\item \textit{resolution:} $\{BC, R \}\not\models \False$
\item \textit{similarity:} $M(G, R) < \epsilon$ 
\end{itemize}
\end{definition}
Intuitively, the first condition states that the refined goals in $R$ remain consistent within the domain properties $Dom$. 
The second condition states that $BC$ does not lead to a divergence situation in the resolution $R$ (i.e., refined goals in $R$ know exactly how to deal with the situations captured by $BC$).  
Finally, the last condition aims at using a similarity metric $M$ to control for the degree of changes applied to the original formulation of goals in $G$  to produce the refined goals in resolution $R$. 

Notice that the similarity metric $M$ is general enough to capture similarities between $G$ and $R$ of  different natures. 
For instance, $M(G, R)$ may compute the \emph{syntactic similarity} between the text representations of the original specification of goals in $G$ and the candidate resolution $R$, where the number of tokens edited from $G$ to $R$ is the aim.
On the other hand, $M(G, R)$ may compute a semantic similarity between $G$ and $R$, for instance, to favour resolutions that weaken the goals (i.e. $G \Implies R$), or strengthen the goals (i.e. $R \Implies G$) or that maintain most of the original behaviours (i.e. $\#G - \#R < \epsilon$). 



Precisely, {\OurTool} will explore syntactic modifications of goals from $G$, leading to newly refined goals in $R$, with the aim at producing candidate resolutions that are consistent with the domain properties $Dom$ and resolve conflict $BC$. 
Assuming that the engineer is competent and the current specification is very close to the intended one~\cite{DeMilloLS78,AcreeBDLS79}, {\OurTool} will integrate two similarity metrics in a multi-objective search process to produce resolutions that are syntactically and semantically similar to the original specification. 
Particularly, {\OurTool} can generate exactly the same resolutions for the MPC previously discussed, manually developed by Letier et al.~\cite{Letier2001}.

\section{{\OurTool}: Automated Goal-Conflict Resolution}
\label{sec:approach}
\begin{figure}[htp!]
\centering
\includegraphics[width=\textwidth]{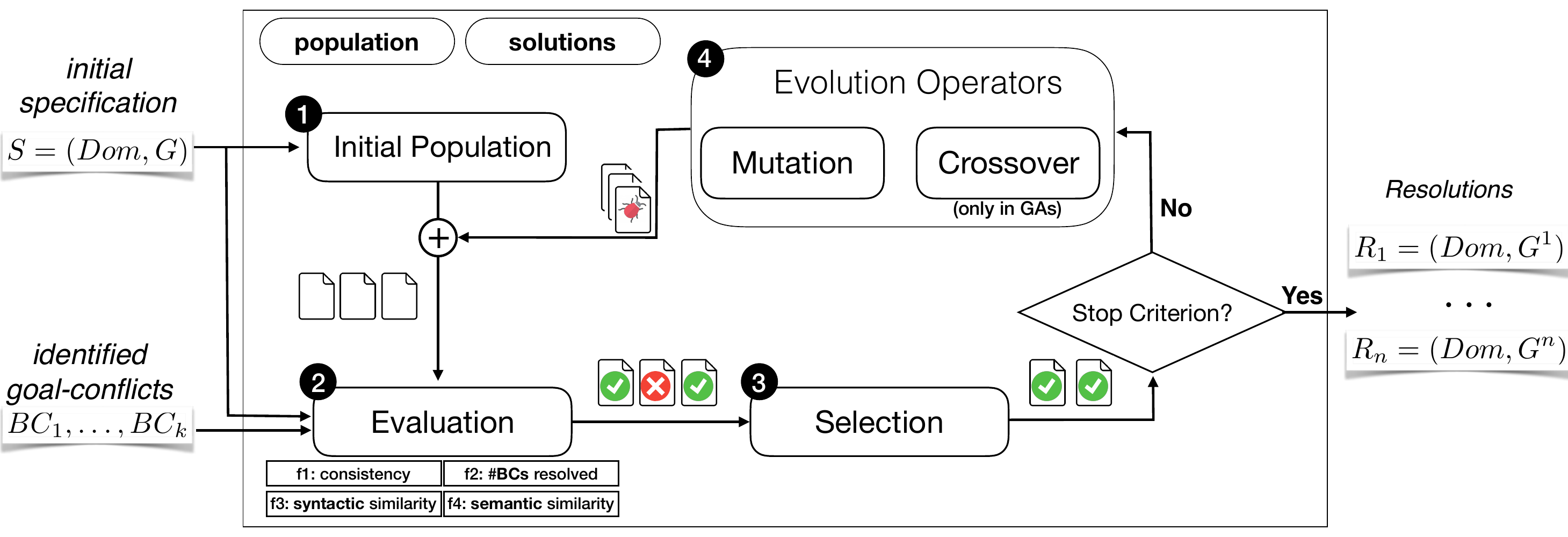}
\vspace{-1.4em}
\caption{Overview of {\OurTool}. }
\label{fig:approach}
\end{figure}

{\OurTool} takes as input a specification $S=(Dom, G)$, composed by the domain properties $Dom$, a set of goals $G$, and a set $\{BC_1,\ldots,BC_k\}$ of identified boundary conditions for $S$. 
{\OurTool} uses search to iteratively explore variants of $G$ to produce a set $R = \{R_1, \ldots, R_n\}$ of resolutions, where each $R_i = (Dom, G^i)$, that maintain two sorts of similarities with the original specification, namely, syntactic and semantic similarity between $S$ and each $R_i$. 
Figure~\ref{fig:approach} shows an overview of the different steps of the search process implemented by {\OurTool}. 

{\OurTool} instantiates multi-objective optimization (MOO) algorithms to efficiently and effectively explore the search space. 
Currently, {\OurTool} implements four MOO algorithms, namely, the \emph{Non-Dominated Sorting Genetic Algorithm III} (NSGA-III)~\cite{DebJain2014}, a \emph{Weight-based genetic algorithm (WBGA)}~\cite{DBLP:books/mit/H1992}, an \emph{Archived Multi-objective Simulated Annealing (AMOSA)}~\cite{DBLP:journals/tec/BandyopadhyaySMD08} approach, and an \emph{unguided search} approach we use as a baseline. 
Let us first describe some common components shared by the algorithms (namely, the search space, the multi-objectives, and the evolutionary operators) and then get into the particular details of each approach (such as the fitness function and selection criteria).

\subsection{Search Space and Initial Population} 
\label{sec:search-space} 
Each individual $cR = (Dom, G')$, representing a candidate resolution, is a LTL specification over a set $AP$ of propositional variables,  where $Dom$ captures the domain properties and $G'$ the refined system goals. 
Notice that domain properties $Dom$ are not changed through the search process since these are descriptive statements. 
On the other hand,  {\OurTool} performs syntactic alterations to the original set of goals $G$ to obtain the new set of refined goals $G'$ that potentially resolve the conflicts given as input.

The initial population represents a sample of the search space from which the search starts. 
{\OurTool} creates one or more individuals (depending on the multi-objective algorithm being used) as the initial population by applying the mutation operator (explained below) to the specification $S$ given as input. 

\subsection{Multi-Objectives: Consistency, Resolution and Similarities}
\label{sec:objectives} 
{\OurTool} guides the search with \emph{four objectives} that check for the validity of each of the conditions needed to be a valid goal-conflict resolution, namely, consistency, resolution and two similarity metrics (cf. Definition~\ref{def:conflict-resolution}). 

Given a resolution $cR = (Dom, G')$, the first objective $\textit{Consistency}(cR)$ evaluates if the refined goals $G'$ are consistent with the domain properties by using SAT solving. 
\begin{equation*}
Consistency(cR) =  
\begin{cases} 
    1 &\mbox{if $Dom \land G'$ is satisfiable}\\
    0.5 &\mbox{if $Dom \land G'$ \blue{is} unsatisfiable, but $G'$ is satisfiable}\\ 
    0 &\mbox{if $G'$ is unsatisfiable}\\ 
    \end{cases}
\end{equation*}

The second objective $\textit{ResolvedBCs}(cR)$ computes the ratio of boundary conditions resolved by the candidate resolution $cR$, among the total number of boundary conditions given as input. 
Hence, $\textit{ResolvedBCs}(cR)$ returns values between $0$ and $1$, and is defined as follows:

\begin{equation*}
    \textit{ResolvedBCs}(cR)= \frac{\sum_{i=1}^{k} isResolved(BC_i, G')}{k} \\
\end{equation*}
$isResolved(cR,BC_i)$ returns $1$, if and only if $BC_i \land G'$ is satisfiable; otherwise, returns $0$. Intuitively, when $BC_i \land G'$ is satisfiable, it means that the refined goals $G'$ satisfies the resolution condition of Definition~\ref{def:conflict-resolution} and thus, $BC_i$ is no longer a conflict for candidate resolution $cR$. 
In the case that $cR$ resolves all the ($k$) boundary conditions, the objective $\textit{ResolvedBCs}(cR)$ will return $1$.  

With the objective of prioritising resolutions that are in some sense similar to the original specification among the dissimilar ones, 
{\OurTool} integrates two similarity metrics. {\OurTool} considers one \emph{syntactic} and one \emph{semantic} similarity metric that will help the algorithms to focus the search in the vicinity of the specification given as input.




Precisely, objective $Syntactic(S,cR)$ refers to the distance between the text representations of the original specification $S$ and the candidate resolution $cR$. 
To compute the syntactic similarity between LTL specifications, we use Levenshtein distance~\cite{10.1145/3300148}. 
Intuitively, the Levenshtein distance between two words is the minimum number of single-character edits (insertions, deletions, or substitutions) required to change one word into the other. 
Hence, $Syntactic(S, cR)$, is computed as:
 $$Syntactic(S, cR) = \frac{maxLength - Levenshtein(S, cR)}{maxLength}$$
where $maxLength = max (length(S), length(cR))$. 
Intuitively, $Syntactic(S, cR)$ represents the ratio between the number of tokens changed from $S$ to obtain $cR$ among the maximum number of tokens corresponding to the largest specification.

On the other hand, our semantic similarity objective $Semantic(S,cR)$ refers to the system \emph{behaviour} similarities described by the original specification and the candidate resolution. 
Precisely, $Semantic(S,cR)$ computes the ratio between the number of behaviours present in both, the original specification and candidate resolution, among the total number of behaviours described by the specifications. 
To efficiently compute the objective $Semantic(S,cR)$, {\OurTool} uses model counting and the approximation previously described in Definition~\ref{def:approxMC}. 
Hence, given a bound $k$ for the lasso traces, the semantic similarity between $S$ and $cR$ is computed as:
\begin{align*}
Semantic(S, cR) = \dfrac{\ApproxMC(S \land cR, k)}{\ApproxMC( S \lor cR, k)}
\end{align*}

Notice that, small values for $Semantic(S, cR)$ indicate that the behaviours described by $S$ are divergent from those described by $cR$. 
In particular, in cases that $S$ and $cR$ are contradictory (i.e., $S \land cR$ is unsatisfiable), $Semantic(S, cR)$ is $0$.  
As this value gets closer to $1$, both specifications characterize an increasingly large number of common behaviors. 

\subsection{Evolutionary Operators} 
New individuals are generated through the application of the evolution operators.  
Particularly, our approach {\OurTool} implements two standard operators used for evolving LTL specifications~\cite{Degiovanni+2018b,LuoWSYZC21}, namely a mutation and a crossover operators. Below, we provide some examples of the application of these operators, and please refer to the complementary material for a detailed formal definition.

\begin{figure}[htp!]
\centering
\begin{minipage}[c]{.35\textwidth}
\centering
\includegraphics[width=.85\textwidth]{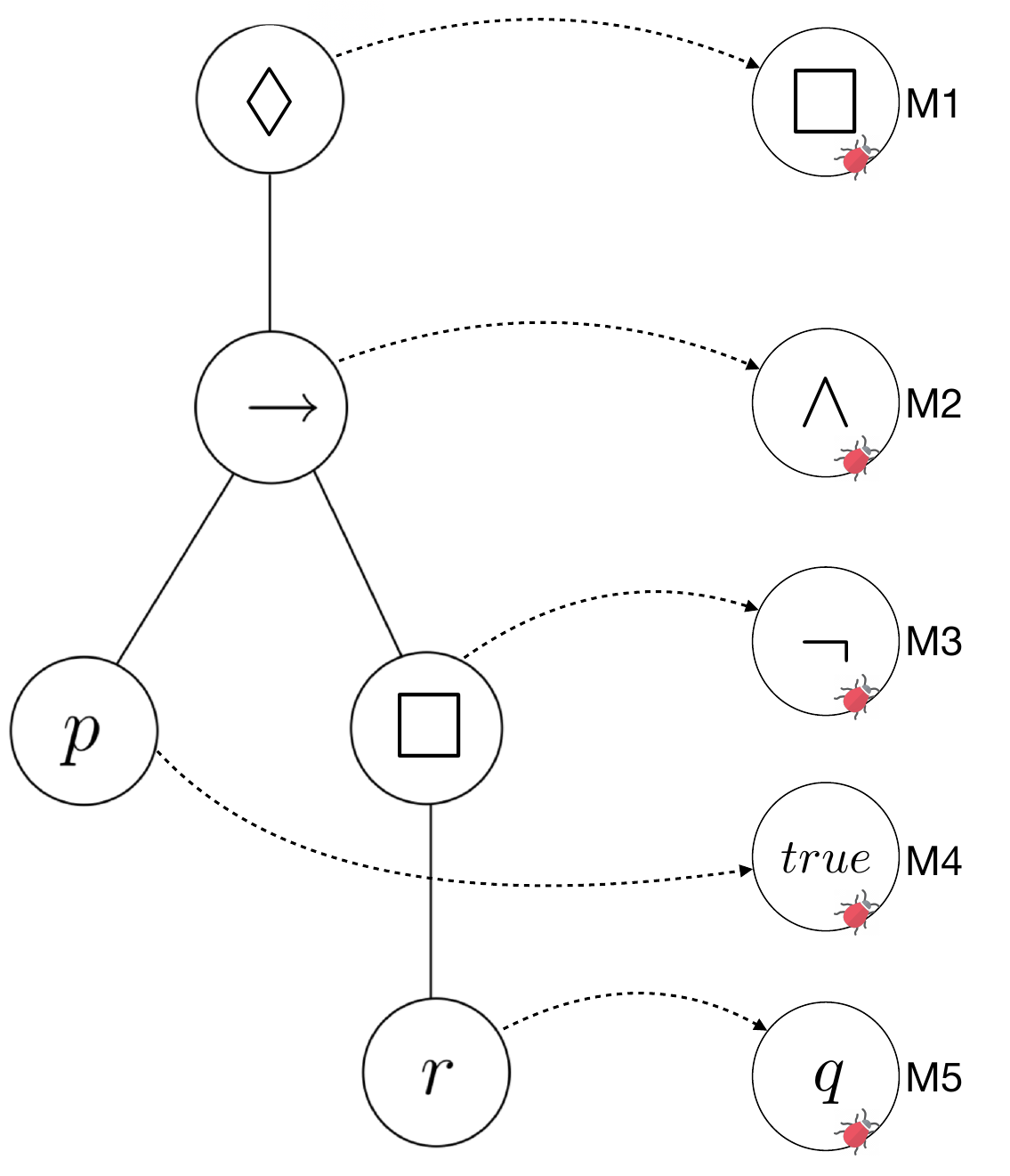}
\caption{Mutation operator.}
\label{fig:mutants}
\end{minipage}
\hfill
\begin{minipage}[c]{.6\textwidth}
\centering
\includegraphics[width=\textwidth]{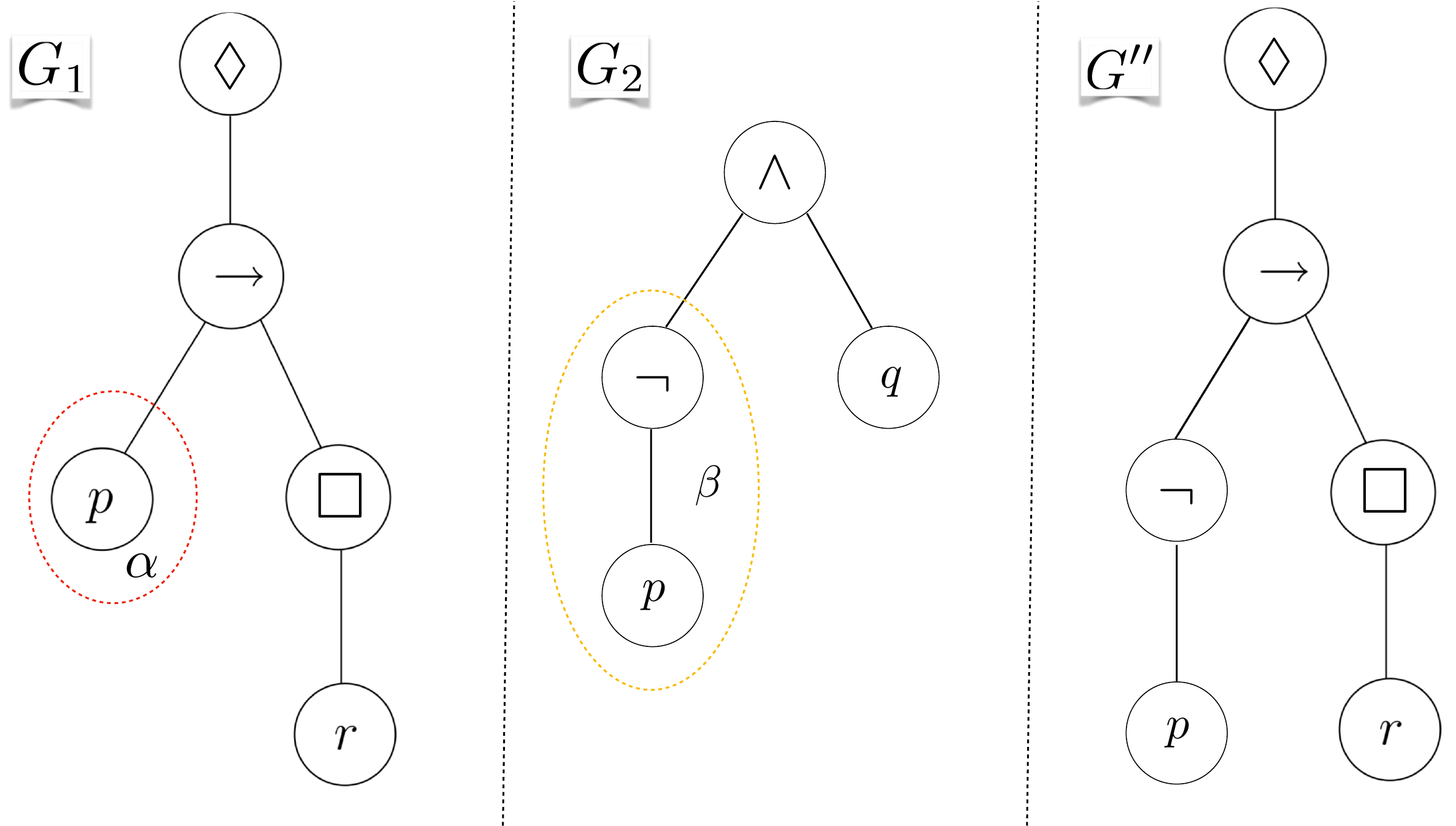}
\vspace{-1.4em}
\caption{Crossover operator.}
\label{fig:crossover}
\end{minipage}
\end{figure}

Given a candidate individual $cR' = (Dom, G')$, the \emph{mutation} operator selects a goal $g' \in G'$ to mutate, leading to a new goal $g''$,  and produces a new candidate specification $cR'' = (Dom, G'')$, where $G''= G' [g' \mapsto g'']$, that is, $G''$ looks exactly as $G'$ but goal $g'$ is replaced by the mutated goal $g''$. 

For instance, Figure~\ref{fig:mutants} shows 5 possible mutations that we can generate for formula $\F (p \Implies \G r)$. 
Mutation M1 replaces $\F$ by $\G$, leading to $M1:\G (p \Implies \G r)$.
Mutation $M2:\F (p \land \G r)$ replaces $\Implies$ by $\land$. 
Mutation $M3:\F (p \Implies \neg r)$ replaces $\G$ by $\neg$.
Mutation $M4: \F (true \Implies \G r)$, reduces to $\F \G r$,  replaces $p$ by $\True$. While mutation $M5 : \F (p \Implies \G q)$ replaces $r$ by $q$.

On the contrary, the \emph{crossover} operator takes two individuals $cR^1 = (Dom, G^1)$ and $cR^2 = (Dom, G^2)$, and produces a new candidate resolution $cR'' = (Dom, G'')$ by combining portions of both specifications. 
In other words, it takes one goal from each individual, i.e. $G_1 \in G^1$ and $G_2 \in G^2$, and generates a new goal $G''$ that is obtained by replacing a subformula $\alpha$ of $G_1$ by a subformula $\beta$ taken from $G_2$.
For instance, Figure~\ref{fig:crossover} provides an illustration of how this operator works. 
Particularly, subformula $\alpha: p$ is selected from goal $G_1:\F (p \Implies \G r)$, while subformula  $\beta:\neg p$ is selected from goal $G_2:\neg p \land q$. 
Hence, by replacing in $G_1$ subformula $\alpha$ by subformula $\beta$, the crossover operators generate a new goal $G'': \F (\neg p \Implies \G r)$.

It is worth mentioning that the four multi-objective search algorithms implemented by {\OurTool} use the mutation operator to evolve the population. 
However, only two of the algorithms that implement two different genetic algorithms (i.e. NSGA-III and WBGA) use the crossover operator to evolve the population.  

\subsection{Multi-Objective Optimisation Search Algorithms}
In a \emph{multi-objective optimisation} (MOO) problem there is a set of solutions, called the \emph{Pareto-optimal} (PO) set, which is considered to be equally important. 
Given two individuals $x_1$ and $x_2$ from the search-space $S$, and $f_1, \ldots, f_n$ a set of (maximising) fitness functions, where $f_i: S \rightarrow \mathbb{R}$, we say that $x_1$ dominates $x_2$ if (a) $x_1$ is not worse than $x_2$ in all objectives and (b) $x_1$ is strictly better than $x_2$ at least in one objective. 
Typically, MOO algorithms evolve the candidate population with the aim to converge to a set of \emph{non-dominated} solutions as close to the true PO set as possible and maintain as diverse a solution set as possible. 
There are many variants of MOO algorithms that have been successfully applied in practice~\cite{Harman:2012:SSE:2379776.2379787}.
{\OurTool} implements four multi-objective optimization algorithms to explore the search space to generate goal-conflict resolutions. 

\noindent
\textbf{AMOSA.} 
The Archived Multi-objective Simulated Annealing (AMOSA)~\cite{DBLP:journals/tec/BandyopadhyaySMD08} is an adaptation of the simulated annealing algorithm~\cite{KirkpatrickGV1983} for multi-objectives. 
AMOSA only analyses one (current) individual per iteration, and a new individual is created by the application of the mutation operator. 
AMOSA has two particular features that make it promising for our purpose. 
During the search, it maintains an ``archive'' with the non-dominated candidates explored so far, that is, candidates whose fitness values are not subsumed by other generated individuals. 
Moreover, when a new individual is created that does not dominate the current one, it is not immediately discarded and can still be selected among the current individual with some probability that depends on the ``temperature'' (a function that decreases over time). At the beginning the temperature is high, then new individuals with worse fitness than the current element, are likely to be selected, but this probability decreases over the iterations. This strategy helps in avoiding local maximums and exploring more diverse potential solutions.


\noindent \textbf{WBGA.} 
{\OurTool} also implements a classic \emph{Weight-based genetic algorithm (WBGA)}~\cite{DBLP:books/mit/H1992}. 
In this case, WBGA maintains a fixed number of individuals in each iteration (a configurable parameter), and applies both the mutation and crossover operators to generate new individuals. 
WBGA computes the fitness value for each objective and combines them into a single fitness $f$ defined as:
\begin{align*}
f(S, cR) = & \alpha * Consistency(cR) + \beta * ResolvedBCs(cR) + \\
& \gamma * Syntax(S, cR) + \delta * Semantic(S, cR)
\end{align*}
\noindent
where weights $\alpha = 0.1 $, $\beta = 0.7$, $\gamma = 0.1$, and $\delta = 0.1$ are defined by default (empirically validated), but these can be configured to other values if desired. 
In each iteration, WBGA sorts all the individuals according to their fitness value (descending order) and selects best ranked individuals to survive to the next iteration (other selectors can be integrated). 
Finally, WBGA reports all the resolutions found during the search. 

\noindent \textbf{NSGA-III.} 
{\OurTool} also implements the Non-Dominated Sorting Genetic Algorithm III (NSGA-III)~\cite{DebJain2014} approach. 
It is a variant of a genetic algorithm that also uses mutation and crossover operators to evolve the population. 
In each iteration, it computes the fitness values for each individual and sorts the population according to the Pareto dominance relation. 
Then it creates a partition of the population according the level of the individuals in the Pareto dominance relation (i.e., non-dominated individuals are in Level-1, Level-2 contains the individuals dominated only by individuals in Level-1, and so on). Thus, NSGA-III selects only one individual per non-dominated level with the aim of diversifying the exploration and reducing the number of resolutions in the final Pareto-front. 

{\OurTool} also implements an \textbf{Unguided Search} algorithm that does not use any of the objectives to guide the search. 
It randomly selects individuals and applies the mutation operator to evolve the population. 
After generating a maximum number of individuals (a given parameter of the algorithm), it checks which ones constitute a valid resolution for the goal-conflicts given as input. 

\section{Experimental Evaluation}
\label{sec:research-questions}
We start our analysis by investigating the effectiveness of {\OurTool} in resolving goal-conflicts. Thus, we ask:

\begin{description}
\item[\textbf{RQ1}] \emph{\red{How effective is} {\OurTool} at resolving goal-conflicts?} 
\end{description}

To answer this question, we study the ability of {\OurTool} to generate resolutions in a set of 25 specifications for which we have identified goal-conflicts. 

Then, we turn our attention to the \textit{``quality''} of the resolution produced by {\OurTool} and study if {\OurTool} is able to replicate some of the manually written resolutions gathered from the literature (ground-truth). Thus, we ask:

\begin{description}
\item[\textbf{RQ2}] \emph{\red{How able is {\OurTool} to} generate resolutions that match with resolutions provided by engineers (i.e. manually developed)?}
\end{description}

To answer RQ2, we check if {\OurTool} can generate resolutions that are equivalent 
to the ones manually developed by the engineer. 

Finally, we are interested in analyzing and comparing the performance of the four search algorithms integrated by {\OurTool}. 
Thus, we ask:
\begin{description}
\item[\textbf{RQ3}] \emph{What is the performance of {\OurTool} when adopting different search algorithms?}
\end{description}
To answer RQ3, we basically employ standard quality indicators (e.g. hypervolume (HV) and inverted generational distance (IGD)) to compare the Pareto-front produced by {\OurTool} when the different search algorithms are employed.


\subsection{Experimental Procedure}
\label{sec:experimental-procedure}
We consider a total of 25 requirements specifications taken from the literature and different benchmarks. These specifications were previously used by goal-conflicts identification and assessment approaches~\cite{Alur+2013,Degiovanni+2018,Degiovanni+2018b,Degiovanni+2016,LuoWSYZC21,vanLamsweerde1998}. 


\begin{table}
\caption{LTL Requirements Specifications and Goal-conflicts Identified.}
\label{tab:subjects}
\begin{minipage}{0.45\textwidth}
\resizebox{.98\textwidth}{!}{
\begin{tabular}{l c c}
\textbf{Specification} & \textbf{$\#Dom + \#Goals$} & \textbf{$\#BCs$}\\
minepump & 3 & 14\\ \hline
simple arbiter-v1 & 4 & 28\\ \hline
simple arbiter-v2 & 4 & 20\\ \hline
prioritized arbiter & 7 & 11\\ \hline
arbiter & 3 & 20\\ \hline
detector & 2 & 15\\ \hline 
ltl2dba27 & 1 & 11\\ \hline
round robin & 9 & 12\\ \hline
tcp & 2 & 11\\ \hline
atm & 3 & 24\\ \hline
telephone & 5 & 4\\ \hline
elevator & 2 & 3\\ \hline
\end{tabular}
}
\end{minipage} \hfill
\begin{minipage}{0.45\textwidth}
\resizebox{.98\textwidth}{!}{
\begin{tabular}{l c c}
\textbf{Specification} & \textbf{$\#Dom + \#Goals$} & \textbf{$\#BCs$}\\
rrcs & 4 & 14\\ \hline
achieve-avoid pattern & 3 & 16\\ \hline
retraction pattern-1 & 2 &  2\\ \hline
retraction pattern-2 & 2 & 10\\ \hline
RG2 & 2 & 9\\ \hline
lily01 & 3 & 5\\ \hline
lily02 & 3 & 11\\ \hline
lily11 & 3 & 5\\ \hline
lily15 & 3 & 19\\ \hline
lily16 & 6 & 38\\ \hline
ltl2dba theta-2 & 1 & 3\\ \hline
ltl2dba R-2 & 1 & 5\\\hline
simple arbiter icse2018 & 11 & 20\\  \hline
\end{tabular}
}
\end{minipage} \hfill
\end{table}

We start by running the approach of Degiovanni et al.~\cite{Degiovanni+2018b} on each subject to identify a set of boundary conditions. 
Table~\ref{tab:subjects} summarises, for each case, the number of domain properties and goals, and the number of boundary conditions (i.e. goal-conflicts) computed with the approach of Degiovanni et al.~\cite{Degiovanni+2018b}. 
Notice that we use the set of ``weakest''\footnote{\label{footnote:weakes-BCs}A formula $A$ is weaker than $B$, if $B \land \neg A$ is unsatisfiable, i.e., if $B$ implies $A$.} boundary conditions returned by \cite{Degiovanni+2018b}, in the sense that by removing all of these we are guaranteed to remove all the boundary conditions computed. 

Then, we run {\OurTool} to generate resolutions that remove all the identified goal-conflicts. 
We configure {\OurTool} to explore a maximum number of 1000 individuals with each algorithm. 
We repeat this process 10 times to reduce potential threats~\cite{RandomizedArcuri} raised by the random elections of the search algorithms. 


To answer RQ1, we run {\OurTool} and report the number of \emph{non-dominated} resolutions produced by each implemented algorithm (i.e. those resolutions whose fitness values are not subsumed by other individuals).  

To answer RQ2, we collected from the literature 8 cases in which authors reported a ``buggy'' version of the specification and a ``fixed'' version of the same specification. 
We take the buggy version and compute a set of boundary conditions for it that are later fed into  {\OurTool} to automatically produce a set of resolutions. 
We then compare the resolutions produced by our {\OurTool} and the ``fixed'' versions we gathered from the literature.  
We basically analyse, by using sat solving, if any of the resolutions produced by {\OurTool} is equivalent to the manually developed fixed version. 


To answer RQ3, we perform an objective comparison of the performance of the four search algorithms implemented by {\OurTool} by using two standard quality indicators: hypervolume (HV)~\cite{Zitzler02performanceassessment} and inverted generational distance (IGD)~\cite{10.1007/978-3-540-24694-7_71}. 
The recent work of Wu et al.~\cite{WuAYAZ2022} indicates that quality indicators HV and IGD are the prefered ones for assessing genetic algorithms and Pareto evolutionary algorithms such as the ones {\OurTool} implements (NSGA-III, WBGA, and AMOSA). 
These quality indicators are useful to measure the \emph{convergence}, \emph{spread}, \emph{uniformity}, and \emph{cardinality} of the solutions computed by the algorithms. 
More precisely, hypervolume (HV)~\cite{10.1145/3300148,DBLP:journals/tec/TanabeI20b} is a \textit{volume-based indicator}, defined by the Nadir Point~\cite{LASZCZYK2019109,Zitzler02performanceassessment}, that
returns a value between 0 and 1, where a value near to 1 indicates that the Pareto-front converges very well to the reference point~\cite{10.1145/3300148} (also, high values for HV are good indicator of uniformity and spread of the Pareto-front~\cite{DBLP:journals/tec/TanabeI20b}). 
The Inverted Generational Distance (IGD) indicator is a \textit{distance-based indicator} that also computes convergence and spread~\cite{10.1145/3300148,DBLP:journals/tec/TanabeI20b}.
In summary, IGD measures the mean distance from each reference point to the nearest element in the Pareto-optimal set~\cite{10.1007/978-3-540-24694-7_71,DBLP:journals/tec/TanabeI20b}. 
We also perform some statistical analysis, namely, the Kruskal-Wallis H-test~\cite{kruskal1952use},  the Mann-Whitney U-test~\cite{MannWhitneyUTest},
and Vargha-Delaney A measure $\hat{A}_{12}$~\cite{VarghaDelaney2000}, to compare the performance of the algorithms. 
Intuitively, the $\textit{p-value}$ will tell us if the performance between the algorithms measured in terms of the HV and IGD is statistical significance, while the A-measure will tell us how frequent one algorithm obtains better indicators than the others.


{\OurTool} is implemented in Java into the JMetal framework
~\cite{10.1145/2739482.2768462}. 
It integrates the LTL satisfiability checker Polsat~\cite{DBLP:journals/corr/LiP0YVH13}, a portfolio tool that runs in parallel with four LTL solvers, helping us to efficiently compute the fitness functions. Moreover, {\OurTool} uses the OwL library~\cite{KretinskyMS18} to parse and manipulate the LTL specifications. The quality indicators also are implemented by the JMetal framework and the statistical tests by the Apache Common Math. 
We ran all the experiments on a cluster with nodes with Xeon E5 2.4GHz, with 5 CPUs-nodes and 8GB of RAM available per run. 

Regarding the setting of the algorithms, the population size of 100 individuals was defined and the fitness evaluation was limited to a number of 1000 individuals. Moreover, the timeout of the model counting and SAT solvers were configured as 300 seconds. The probability of crossover application was 0.1, while mutation operators were always applied. A tournament selection of four solutions was used for NSGA-III, while WBGA instantiated Bolzman’s selection with a decrement exponential function. The WBGA was configured to weight the fitness functions as a proportion of 0.1 in the Status, 0.7 in the ResolvedBC, 0.1 in Syntactic, and 0.1 in Semantic. The AMOSA used an archive of crowding distance, while the cooling scheme relied on a decrement exponential function.

The case studies and results are publicly available at \url{https://sites.google.com/view/acore-goal-conflict-resolution/}.

\section{Experimental Results}
\label{sec:experimental-results}

\subsection{\blue{RQ1}: Effectiveness of {\OurTool} }

Table~\ref{table:rq1} reports the average number of non-dominated resolutions produced by the algorithms in the 10 runs.  
First, it is worth mentioning that when {\OurTool} uses any of the genetic algorithms (NSGA-III or WBGA), it successfully generates at least one resolution for all the case studies. 
However, AMOSA fails in producing a resolution for the \texttt{lily16} and \texttt{simple arbiter icse2018} in 2 and 1 cases
of the 10 runs, respectively. Despite that Unguided search \red{succeeds} in the majority of the cases, it was not able to produce any resolution for the \texttt{prioritized arbiter}, and failed in producing a resolution in 5 out of the 10 runs for the \texttt{simple-arbiter-v2}.

\begin{table}
\begin{minipage}[b]{0.57\textwidth}
\caption{Effectiveness of {\OurTool} in producing resolutions.}
\label{table:rq1}
\resizebox{\textwidth}{!}{
\begin{tabular}{l r r r r}
\toprule
\textbf{Specification}
& \multicolumn{1}{c}{\textbf{NSGA-III}}
& \multicolumn{1}{c}{\textbf{WBGA}}
& \multicolumn{1}{c}{\textbf{AMOSA}} 
& \multicolumn{1}{c}{\textbf{Unguided}}  \\
\midrule
minepump                & 5.0           & \textbf{6.5}   & 1.8  & 5.1 \\ \hline
simple arbiter-v1       & \textbf{4.8}  & 3.1            & 2.0   & 4.1 \\ \hline
simple arbiter-v2       & 3.1           & \textbf{3.4}   & 2.3   & 0.5 \\ \hline
prioritized arbiter     & 3.1           & \textbf{3.7}   & 2.2    & 0.0 \\ \hline
arbiter                 & \textbf{5.8}  & 2.7            & 3.0   & 5.5 \\ \hline
detector                & 4.9           & 4.8            & 3.2   & \textbf{6.1} \\ \hline
ltl2dba27               & 3.0           & \textbf{4.2}   & 3.5   & 4.0 \\ \hline 
round robin             & \textbf{7.0}  & 4.2            & 4.7   & 4.7 \\ \hline
tcp                     & 6.4           & 4.9            & 2.0  & \textbf{7.4} \\ \hline
atm                     & 3.9           & \textbf{6.3}   & 3.3    & 4.5 \\ \hline
telephone               & \textbf{4.7}  & 4.4            & 2.2   & 4.5 \\ \hline
elevator                & \textbf{5.9}  & \textbf{5.9}   & 3.6   & 4.8 \\ \hline
rrcs                    & 5.5           & \textbf{5.7}   & 1.4    & 3.3 \\ \hline
achieve pattern         & 5.0           & \textbf{5.9}   & 2.5   & 2.8 \\ \hline
retraction pattern-1    & 4.1           & 4.0            & 2.7   & \textbf{4.6} \\ \hline
retraction pattern-2    & \textbf{6.1}  & 4.8            & 2.6   & 6.0 \\ \hline
RG2                     & 3.3           & \textbf{5.2}   & 1.5    & 4.3 \\ \hline
lily01                  & \textbf{5.1}  & \textbf{5.1}   & 1.5   & 4.1 \\ \hline
lily02                  & 2.4           & \textbf{3.8}   & 1.9    & 1.9 \\ \hline
lily11                  &  \textbf{7.1} & 5.0            & 2.2   & 5.8 \\ \hline
lily15                  & \textbf{6.1}  & 4.1            & 1.2   & 5.8 \\ \hline
lily16                  & \textbf{3.5}  & 3.2            & 0.8   & 3.8 \\ \hline
ltl2dba theta-2         & 1.9           & \textbf{2.8}   & 1.9    & 1.2 \\ \hline
ltl2dba R-2             & 1.0           & \textbf{2.1}   & 1.9    & \textbf{2.1} \\ \hline
simple arbiter icse2018 & \textbf{3.8}  & 3.7            & 0.9   & 3.5 \\ 
\bottomrule
\end{tabular}
}
\end{minipage}
\hfill
\begin{minipage}[b]{0.43\textwidth}
\centering
\caption{{\OurTool} effectiveness in producing an exact or more general resolution than the manually written one.} 
\label{tab:comparison-ground-truth}
\resizebox{\textwidth}{!}{
\begin{tabular}{l c c c c}
\toprule
\textbf{Specification}  & \textbf{NSGA-III} & \textbf{WBGA} & \textbf{AMOSA} & \textbf{Unguided}\\
\midrule
minepump                & \checkmark    & \checkmark    & \checkmark   & \checkmark \\ \hline
simple arbiter-v1       &               &               &\\ \hline
simple arbiter-v2       &   \checkmark  & \checkmark    &               & \\ \hline
prioritized arbiter     & \\ \hline
arbiter                 &               & \\ \hline
detector                & \checkmark    & \checkmark    &               & \checkmark \\ \hline 
ltl2dba27               &               & \\ \hline
round robin             &               & \\ 
\bottomrule
\end{tabular}
}
\end{minipage}
\end{table}

Second, the genetic algorithms (NSGA-III and WBGA) generate on average more (non-dominated) resolutions than AMOSA and unguided search. 
The results point out that WBGA generates more (non-dominated) resolutions than others in 13 out of the 25 cases, and  NSGA-III is the one that produces more (non-dominated) resolutions in 11 cases. 
Considering the genetic algorithms together, we can observe that they outperform the AMOSA and unguided search in 21 out of the 25 cases, and coincide in one case (\texttt{ltl2dba R-2}). 
Finally, the Unguided Search generates more resolutions in 3 cases, namely, \texttt{detector}, \texttt{TCP}, and \texttt{retraction-pattern-1}. 
Interestingly, the different algorithms of {\OurTool} produce on average between 1 and 8 non-dominated resolutions, which we consider is a reasonable number of options that the engineer can manually inspect and validate to select the most appropriate one. \\

\noindent
\fbox{%
    \parbox{11.90cm}{
        {\OurTool} generates more non-dominated resolutions when adopting genetic algorithms. On average, {\OurTool} produces between 1 and 8 non-dominated resolutions that can be presented to the engineer for analysis and validation.
    }
}

\subsection{RQ2: Comparison with the Ground-truth}
Table~\ref{tab:comparison-ground-truth} presents the effectiveness of {\OurTool} in generating a resolution that is equivalent or more general than the ones manually developed by engineers. 
Overall, {\OurTool} is able to reproduce same resolutions in 3 out of 8 of the cases, namely, for the \texttt{minepump} (our running example), \texttt{simple arbiter-v2}, and \texttt{detector}. 
Like for RQ1, the genetic algorithms outperform AMOSA and unguided search in this respect. Particularly, the Unguided Search can replicate the resolution for the \texttt{detector} case, in which AMOSA fails.





\noindent
\fbox{%
    \parbox{11.90cm}{
        Overall, the genetic algorithms can produce same or more general resolutions than the ground-truth in 3 out of the 8 cases, outperforming AMOSA (1 out of the 8) and unguided search (2 out of the 8).
    }
}

\subsection{RQ3: Comparing the Multi-objective Optimization Algorithms}
For each set of non-dominated resolutions generated by the different algorithms, we compute the quality indicators HV and IGD for the syntactic and semantic similarity values. 
The reference point is the best possible value for each objective which is 1. 
These will allow us to determine which algorithm converges the most to the reference point and produces more diverse and optimal resolutions.

\begin{figure}[htp!]
\centering
\begin{minipage}[b]{0.48\textwidth}
\includegraphics[width=.99\textwidth]{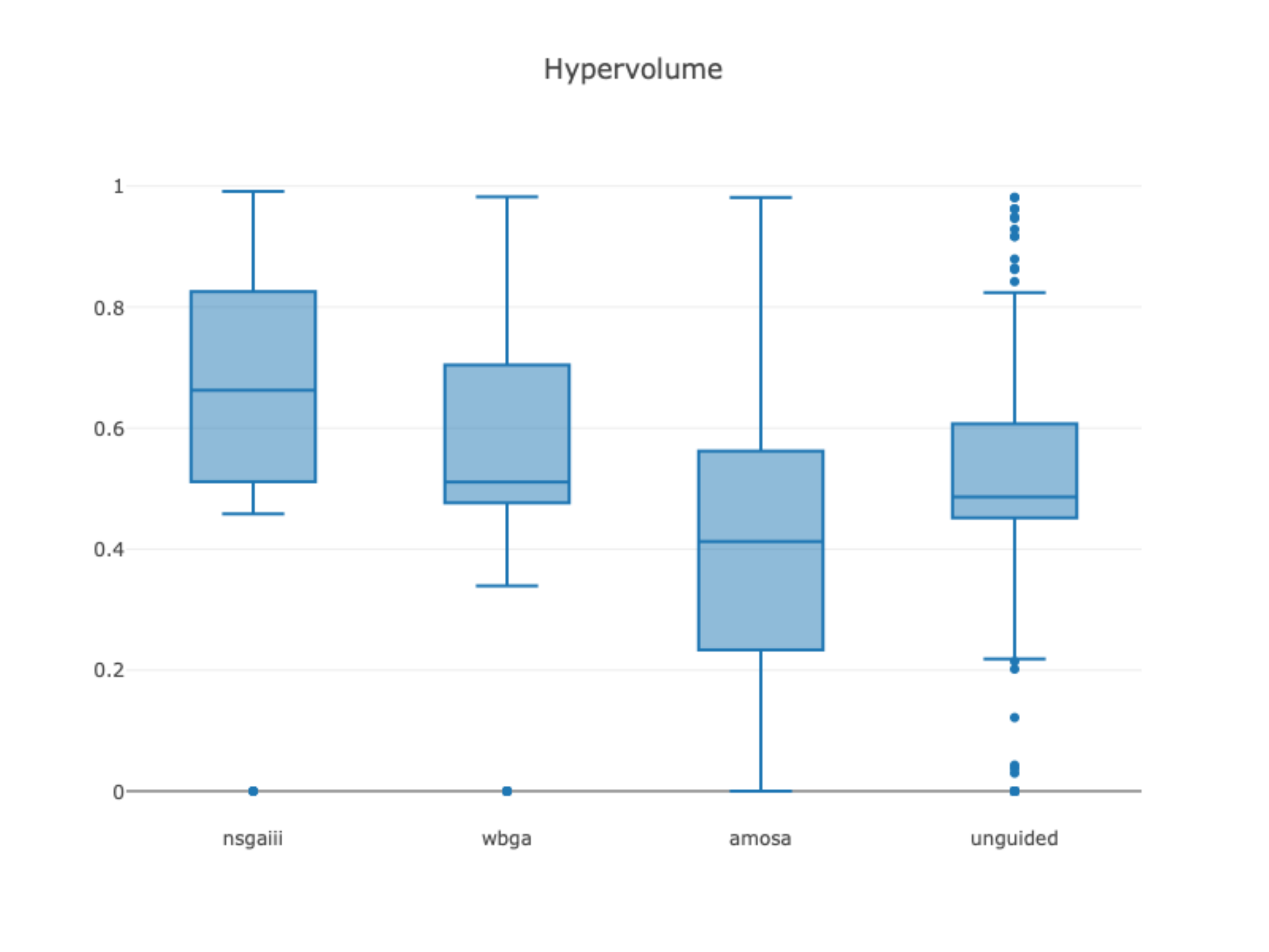}
\caption{\blue{HV of the Pareto-optimal sets generated by {\OurTool}.}}
\label{fig:repairs-HV}
\end{minipage}
\hfill
\begin{minipage}[b]{0.48\textwidth}
\includegraphics[width=.99\textwidth]{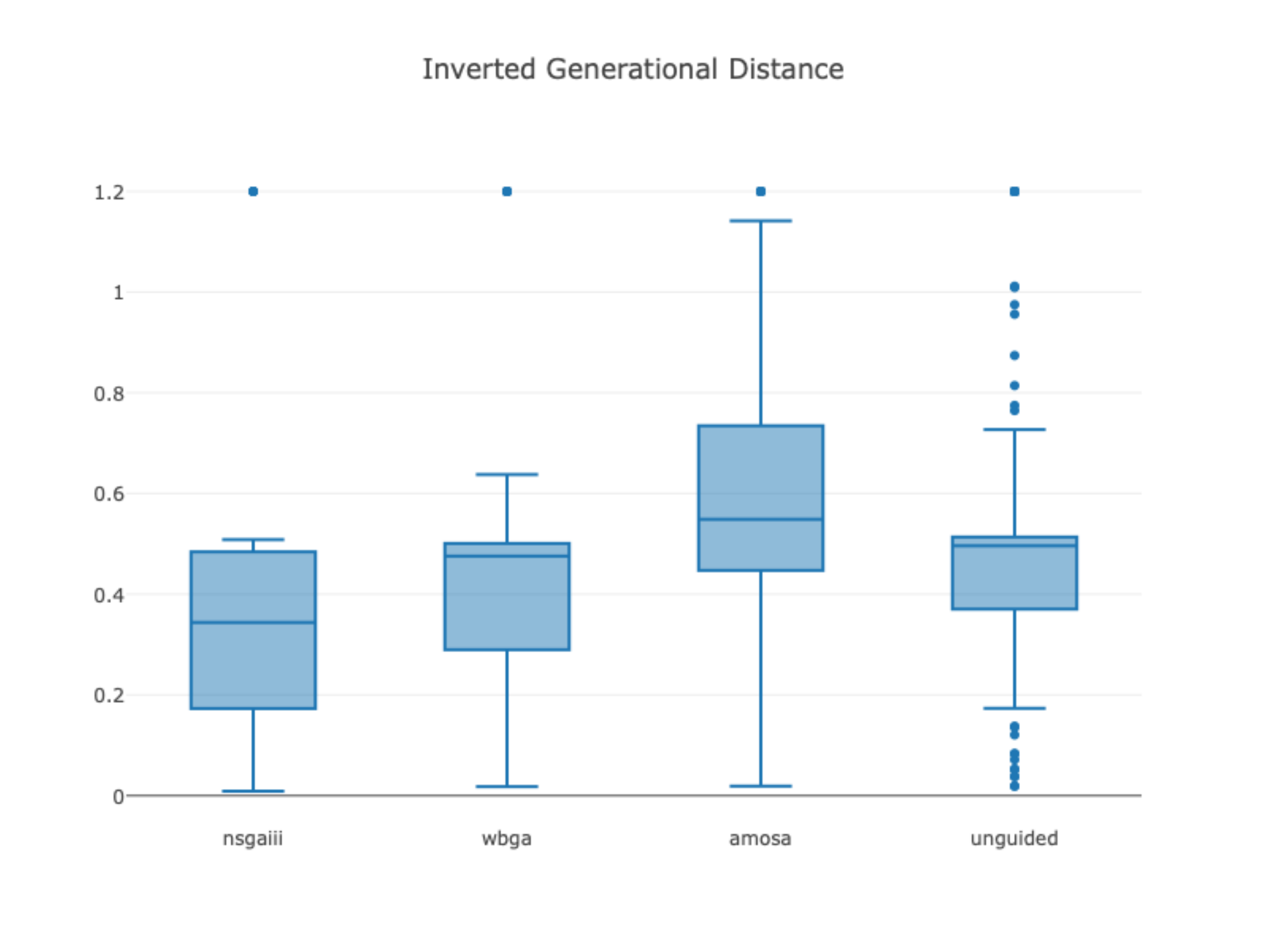}
\caption{\blue{IGD of the Pareto-optimal sets generated by {\OurTool}.}}
\label{fig:repairs-IGD}
\end{minipage}
\end{figure}

Figures~\ref{fig:repairs-HV} and~\ref{fig:repairs-IGD} show the boxplots for each quality indicator. 
NSGA-III obtains on average much better HV and IGD than the rest of the algorithms. 
Precisely, it obtains on average 0.66 of HV (while higher the better) and 0.34 of IGD (while lower the better), outperforming the other algorithms. 

To confirm this result we compare the quality indicators in terms of non-parametric statistical tests: (i) Kruskal–Wallis test by ranks and (ii) the Mann-Whitney U-test. The $\alpha$ value defined in the Kruskal-Wallis test by ranks is $0.05$ and the Mann-Whitney U-test is $0.0125$. Moreover, we also complete our assessment by using Vargha and Delaney's $\hat{A}_{12}$, a non-parametric effect size measurement. 
Table~\ref{tab:metrics-opt} summarises the results when we compare pair-wise each one of the approaches.
We can observe that NSGA-III in near 80\% of the cases obtains resolutions with better quality indicators than AMOSA and Unguided search (and the differences are statistically significant). We can also observe that NSGA-III obtains higher HV (IGD) than WBGA in 66\% (65\%) of the cases. 
From Table~\ref{tab:metrics-opt} we can also observe that WBGA outperforms both AMOSA and unguided search. 
Moreover, we can observe that AMOSA is the worse performing algorithm according to the considered quality indicators. 

\begin{table}[htp!]
\centering
\caption{HV and IGD quality indicators for the generated resolutions.}
\label{tab:metrics-opt}
\resizebox{.80\textwidth}{!}{
\small
\begin{tabular}{ll | rr| rr| rr}
 &
& \multicolumn{2}{c}{\textbf{WBGA}}
& \multicolumn{2}{c}{\textbf{AMOSA}} 
& \multicolumn{2}{c}{\textbf{Unguided}}  \\
& & 
\textbf{$HV$} & \textbf{$IGD$}& \textbf{$HV$} & \textbf{$IGD$}& \textbf{$HV$} & \textbf{$IGD$} \\ 
\multirow{1}{*}{\textbf{NSGAIII}} & \textit{p-value} 
& $<$ 0.00001 & $<$ 0.00001 & $<$ 0.00001  & $<$ 0.00001 & $<$ 0.00001  & $<$ 0.00001 \\ 
& \textit{$\hat{A}_{12}$} 
& 0.66 & 0.65 & 0.84  & 0.83 & 0.80  & 0.76 \\ \hline
\multirow{1}{*}{\textbf{WBGA}} & \textit{p-value} 
& - & - & $<$ 0.00001 & $<$ 0.00001 & $<$ 0.00001 & $<$ 0.00001 \\ 
& \textit{$\hat{A}_{12}$} 
& - & - & 0.74 & 0.74 & 0.64 & 0.61 \\ \hline
\textbf{\textbf{AMOSA}} & \textit{p-value} 
& - & - & - & - & $<$ 0.00001 & $<$ 0.00001\\
& \textit{$\hat{A}_{12}$} 
& - & - & -  & - & 0.36 & 0.36 \\ 
\end{tabular}
}
\end{table}


\noindent
\fbox{%
    \parbox{11.85cm}{
       Overall, both statistical tests evidence that NSGA-III leads to a set of resolutions with better quality indicators (HV and IGD) than the rest of the algorithms. WBGA is the one in the second place, outperforming the unguided search and AMOSA. 
       While AMOSA shows the lowest performance based on the quality indicators, even worse than the unguided search in several cases. 
    }
}

\section{Related Work}
\label{sec:related-work}

Several manual approaches have been proposed to identify inconsistencies between goals and resolve them once the requirements were specified. Among them, Murukannaiah~\textit{et al.}~\cite{Murukannaiah:2015} compares a genuine analysis of competing hypotheses against modified procedures that include requirements engineer thought process. The empirical evaluation shows that the modified version presents higher completeness and coverage. Despite the increase in quality, the approach is limited to manual applicability performed by 
engineers as well previous approaches~\cite{vanLamsweerde1998}. 

Various informal and semi-formal approaches~\cite{Hausmann:2002,Kamalrudin:2009,Kamalrudin:2011}, as well as more formal approaches~\cite{Ellen:2014,Ernst:2012,Harel:2005,Hunter:1998,Nguyen:2013,Spanoudakis:1997}, have been proposed for detecting logically inconsistent requirements, a strong kind of conflicts, as opposed to this work that focuses on a weak form of conflict, called divergences (cf. Section~\ref{sec:GORE}). 


Moreover, recent approaches have been introduced to automatically identify goal-conflicts. Degiovanni~\textit{et al.}~\cite{Degiovanni+2016} introduced an automated approach where boundary conditions are automatically computed using a tableaux-based LTL satisfiability checking procedure. 
Since it exhibits serious scaliability issues, the work of Degiovanni~\textit{et al.}~\cite{Degiovanni+2018b} proposes a genetic algorithm that mutates the LTL formulas in order to find boundary conditions for the goal specifications. 
The output of this approach can be fed into {\OurTool} to produce potential resolutions for the identified conflicts (as shown in the experimental evaluation).

Regarding specification repair approaches, Wang~\textit{et al.}~\cite{8802763} introduced ARepair, an automated tool to repair a faulty model formally specified in Alloy~\cite{Jackson2006}. 
ARepair takes a faulty Alloy model and a set of failing tests and applies mutations to the model until all failing tests become passing. 
In the case of {\OurTool}, 
the identified goal conflicts are the ones that guide the search, and candidates are aimed to be syntactic and semantically similar to the original specification. 

In the context of reactive synthesis~\cite{EmersonClarke1982,MannaWolper1984,PnueliRosner1989}, some approaches were proposed to repair imperfections in the LTL specifications that make the unrealisable ( i.e., no implementation that satisfies the specification can be synthesized). The majority of the approaches focus on learning missing assumptions about the environment that make them unrealisable~\cite{Alur+2013,CavezzaAlrajeh2016,Chatterjee+2008,Maoz+2019}.  
A more recent approach~\cite{DBLP:journals/corr/abs-2105-12595}, published in a technical report, proposes to mutate both the assumptions and guarantees (goals) until the specification becomes realisable.  
Precisely, we use the novel model counting approximation algorithm from Brizzio et. al~\cite{DBLP:journals/corr/abs-2105-12595} to compute the semantic similarity between the original buggy specification and the resolutions. 
However, the notion of repair for Brizzio et. al~\cite{DBLP:journals/corr/abs-2105-12595} requires a realizable specification, which is very general and does not necessarily lead to quality synthesized controllers~\cite{DIppolito+2013,MaozRingert2016b}. 
In this work, the definition of resolution is fine-grained and focused on removing the identified conflicts, which potentially leads to interesting repairs as we showed in our empirical evaluation. 

Alrajeh~\textit{et al.}~\cite{Alrajeh+2020} introduced an automated approach to refine a goal model when the environmental context changes. That is, if the domain properties are changed, then this approach will propose changes in the goals to make them consistent with the new domain. The adapted goal model is generated using a new counterexample-guided learning procedure that ensures the correctness of the updated goal model, preferring more local adaptations and more similar goal models. 
In our work, the domain properties are not changed and the adaptions are made to resolve the identified inconsistencies, and instead of counter-examples, our search is guided by syntactic and semantic similarity metrics. 

\section{Conclusion}
\label{sec:conclusion}


In this paper, we presented {\OurTool}, the first automated approach for goal-conflict resolution. 
Overall, {\OurTool} takes a goal specification and a set of conflicts previously identified, expressed in LTL, and computes a set of resolutions that removes such conflicts. 
To assess and implement {\OurTool} that is a search-based approach, we adopted three multi-objective algorithms (NSGA-III, AMOSA, and WBGA) 
that simultaneously optimize and deal with the trade-off among the objectives. 
We evaluated {\OurTool} in 25 specifications that were written in LTL and extracted from the related literature. 
The evaluation showed that the genetic algorithms (NSGA-III and WBGA) typically generate more (non-dominated) resolutions than AMOSA and an Unguided Search we implemented as a baseline in our evaluation. 
Moreover, the algorithms generate on average between 1 and 8 resolutions per specification, which may allow the engineer to manually inspect and select the most appropriate resolutions. 
We also observed that the genetic algorithms (NSGA-III and WBGA) outperform AMOSA and Unguided Search in terms of several quality indicators: number of (non-dominated) resolutions and standard quality indicators (HV and IGD) for multi-objective algorithms.

\bibliographystyle{plain}
\bibliography{Bibliography}

\end{document}